\documentclass[conference]{IEEEtran}
\IEEEoverridecommandlockouts
\usepackage{cite}
\usepackage{amsmath,amssymb,amsfonts}
\usepackage[utf8]{inputenc}
\usepackage{algorithmic}
\usepackage{graphicx}
\usepackage{textcomp}
\usepackage{xcolor}
\def\BibTeX{{\rm B\kern-.05em{\sc i\kern-.025em b}\kern-.08em
    T\kern-.1667em\lower.7ex\hbox{E}\kern-.125emX}}
\begin{document}

\title{   A Cooperative NOMA User Pairing in UAV-Based Wireless Networks
\\
%{\footnotesize \textsuperscript{*}Note: Sub-titles are not captured in Xplore and
%should not be used}
%\thanks{Identify applicable funding agency here. If none, delete this.}
}

\author{\IEEEauthorblockN{Ghania Khraimech, Fatiha Merazka}
\IEEEauthorblockA{\textit{LISIC Lab.  Telecommunications Department,} \\
\textit{Electrical Engineering Faculty,}\\
\textit{USTHB University,}\\
16111, Algiers, Algeria\\
\{gkhraimech,fmerazka\}@usthb.dz}
%\and
%\IEEEauthorblockN{3\textsuperscript{rd} Given Name Surname}
%\IEEEauthorblockA{\textit{dept. name of organization (of Aff.)} \\
%\textit{name of organization (of Aff.)}\\
%City, Country \\
%email address or ORCID}
%\and
%\IEEEauthorblockN{4\textsuperscript{th} Given Name Surname}
%\IEEEauthorblockA{\textit{dept. name of organization (of Aff.)} \\
%\textit{name of organization (of Aff.)}\\
%City, Country \\
%email address or ORCID}
%\and
%\IEEEauthorblockN{5\textsuperscript{th} Given Name Surname}
%\IEEEauthorblockA{\textit{dept. name of organization (of Aff.)} \\
%\textit{name of organization (of Aff.)}\\
%City, Country \\
%email address or ORCID}
%\and
%\IEEEauthorblockN{6\textsuperscript{th} Given Name Surname}
%\IEEEauthorblockA{\textit{dept. name of organization (of Aff.)} %\\
%\textit{name of organization (of Aff.)}\\
%City, Country \\
%email address or ORCID}
}

\maketitle

\begin{abstract}
NOMA (non-orthogonal multiple access) will be seen as a promising technology for enhancing spectrum efficiency in future cellular networks. The use of unmanned aerial vehicle (UAV) Amplify-and-Forward (AF) relaying as a moving access point or BS, on the other hand, has emerged as a potential solution to wireless networks' high traffic demands. In this paper, we study joint user pair and resource allocation-based distance to optimize fair throughput in a downlink scenario, which concentrates on UAV-aided communication from different wireless-powered nodes. Several transmission methods are proposed, including NOMA as well as UAV cooperative relaying and two representative node-pairing strategies. The simulation results illustrate that the proposed user pairing strategies for cooperative NOMA and UAV-based cellular networks enhance downlink transmission performance and ensure optimum use of power and bandwidth resources.
\end{abstract}


\section{Introduction}
Due to considerable technological developments across several drone-related domains ranging from embedded systems to autonomy, control, security, and communications, the domain of unmanned aerial vehicles (UAVs), sometimes known as drones, has undergone a significant revolution in recent years. UAV communications are also a useful strategy to provide communication links throughout temporary events and after catastrophes in distant regions missing cellular infrastructure \cite{b1}-\cite{b5}. One of the encouraging use case scenarios of UAVs is designing flexible, adaptable, and wireless multiple antennas in the sky \cite{b6}. Because of the high elevation of UAVs, line-of-sight (LoS) in UAV communications can provide better small-scale fading between UAVs and ground customers than in traditional ground links, which presents both opportunities, as well as challenges in the development of UAV cellular networks \, cite{b7}. Due to the obvious low - power consumption of UAVs, achieving enhanced spectrum and energy efficiency is essential to the ability to get the most out of UAV-based network technologies \cite{b8}-\cite{b10}.

NOMA seems to be a useful product for achieving both spectrum efficiency and energy efficiency in next-generation wireless technologies and beyond, especially in UAV communication networks \cite{b11}. The potential advantage of NOMA over OMA in a cellular communication system where a base station (BS) is equipped with enormous antenna arrays was successfully investigated in \cite{b12}. In comparison to conventional OMA techniques, NOMA is capable of a more efficient manner utilizing available resources by arbitrarily maximizing users' specific fading channel on both single-cell and cellular systems \cite{b13},\cite{b14}, and it is ready to accommodate multiple members with specific quality-of-service (QoS) criteria in the same users access \cite{b15}–\cite{b17}. Theoretically, spectral efficiency combining within a frequency, time, and code block delivers the composite signal to multiple customers concurrently in NOMA. NOMA is based on the use of superposition coding (SC) at the transmitter and successive interference cancellation (SIC) techniques at the receiver \cite{b4}, \cite{b12}. Multiple access users in the power domain can be implemented in this scenario by using various energy degrees for consumers in the same resource block. As a result, by employing NOMA techniques to increase the available spectrum and energy efficiency, UAV networks may offer several users at the same time \cite{b18}. 

The downlink of a UAV-enabled wireless communication is investigated in this article using power-domain NOMA and cooperative relaying, with the ground nodes being wireless-powered equipment. Air-to-ground (A2G) communication channels\cite{b19} are used by these devices, and they are characterized by an altitude-dependent route loss exponent and fading. We focus on a user pairing system for NOMA with cooperative relaying cellular networks, in which access devices are separated into two groups and a unit pair is produced from each group. After that, the channel capacity is divided by the number of pairs, with each pair sharing the same sub-channel to deliver their data. The primary issues that we address in this work are related to how to pair NOMA users within each available orthogonal resource, user-pairing strategy for NOMA as well as cooperative relaying in our UAV model, which has a better performance given a specific user pairing strategy.

The remainder of this paper is organized as follows. In Section II, we present the system and channel models considered throughout this article.  Section III details user pairing in cooperative NOMA. Performance analysis of the Cooperative NOMA-based transmission and pairing strategies evaluated in this work are described in Section IV. The simulation results are provided and discussed in Section V. Finally, Section VI presents general conclusions regarding the obtained results in this paper as well as prospects for our future work.

\section{System Model}

\begin{figure}[htbp]
\centerline{\includegraphics[width=8cm,height=7cm]{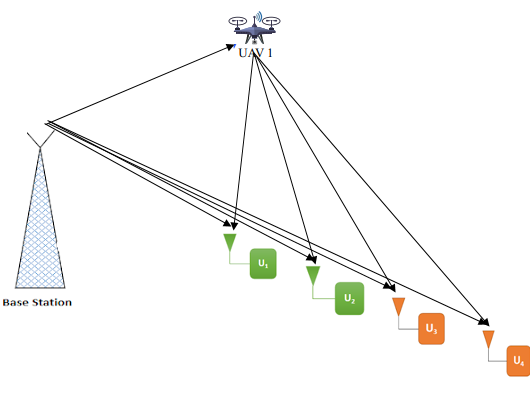}}
\caption{Illustration of the system model.}
\label{fig1}
\end{figure}

As shown in Fig. \ref{fig1}, we consider a link between a transmitter BS and receivers, 4 users, separated by a distance $R_i$. We assume that direct communication between the transmitter and the receivers is possible, where there is one source node BS, $K$ half-duplex amplify-and-forward (AF) UAV relay nodes denoted as $R_k, k = 1, ... , K$ and four users $U_1, U_2, U_3,$ and $U_4$, respectively. We use  UAVs, each equipped with a single antenna, to enable this link and maximize its capacity. Each UAV acts as an amplify-and-forward relay where it simply receives the signal, amplifies, and re-transmits it in a synchronized manner.

 $[H]_1 \in C\textsuperscript{\(N_T\)×\(N_R\)}$ denotes the communication between the transmitter and the UAVs, whereas,  $[H]_2 \in C\textsuperscript{\(N_R\)×\(N_U\)}$  denotes the channel between the UAVs and the receiver. We assume that both channels are strong LoS channels with each element defined as $[H]_ {i,j} =\frac{ \lambda }{4\pi d_{i,j} } e^{\frac{j2 \pi  d_{i,j}}{\lambda}} $, where $ \lambda $ where are the signal frequency band and $d_{i,j}$ is the distance between antenna elements $i$ and $j$ at the transmitter, relay or receiver. 

\section{User Pairing in Cooperative NOMA}
\subsection{Cooperative NOMA}
We know that NOMA exploits successive interference cancellation (SIC) \cite{b20}, in which one user detects the other user's message from the superposition-coded incoming signal before decoding his own. Thus, while conducting SIC, the near user decodes the information of the far user. This is a step that cannot be ignored. In every case, the data of the far user must be decoded by the near user.

Now that the near user has access to the far user's data, they can help him by relaying it to him. Because the distant user's channel with the broadcasting BS is poor, the close user's retransmission of his data will supply him with diversity. In other words, he'll get two copies of the same message. One comes from the BS, and the other comes from a nearby user who is serving as a relay. As a result, we can assume the far user's outage probability to reduce.

Cooperative communication/cooperative relaying \cite{b21} is the name for this concept. We can see that NOMA naturally encourages cooperative communication because the near user has access to the data of the far user, which they must decode anyway. The benefit of cooperative communication is link redundancy, which allows a message to be transmitted twice. In those other words, if one link is down, the other is very certainly up and running. The scenario in which both links fall at the same time is less likely than one link going down at a time. We have a lower outage probability and, as a result, more diversity benefits without the need for extra antennas in this situation (i.e., MIMO). Another benefit of relaying is that it might essentially extend the BS's coverage area. Let's create a cooperative NOMA network now that we've seen what cooperative communication is and how it might benefit our network. Under a line-of-sight channel, we'll consider a downlink transmission with a BS and two NOMA users relayed by a UAV acting as an amplify-and-forward relay.\cite{b22} With the UAV, we have a near user with a stronger channel and a distance user with weak channel circumstances. The transmission is split into two parts. Let us refer to the first time slot as "direct transmission" and the second as "relaying." 
\subsubsection{Direct transmission slot}

In the direct transmission slot, the BS uses NOMA to transmit data intended for the near user \(x_n\) and the far user \(x_f\). The near user uses SIC to decode the far user's data first and then proceeds to decode its data. The far user will only perform direct decoding.

At the end of the direct transmission slot, the achievable data rates at the near user and far user are,

\begin{equation}
R_{n}= \frac{1}{2} log_{2}  ( 1 + \alpha_{n} \rho \lvert h_{n}\rvert ^2).
\end{equation}
for the detection of SIC, the desired signal and interference signal is presented as 

\begin{equation}
R_{f,n}= \frac{1}{2} log_2  ( 1 + \frac{\alpha_f \rho \lvert h_f\rvert ^2}{{\alpha_n} \rho \lvert h_f\rvert ^2 + 1} )\label{eq1}.
\end{equation}

Notation
\begin{itemize}
\item $\alpha_n$: power allocation coefficient for the near user.
\item $\alpha_f$: power allocation coefficient for the far user.
\item $h_n$ : the channel between BS and near the user.
\item $h_f$ : the channel between BS and far user.
\item $\rho$: transmit $SNR =\frac{P}{\sigma^2}$, where $P$ is the transmit power and $\sigma^2$ is the noise variance
\item  As usual, $\alpha_f> \alpha_n$, and, $\alpha_n$+$\alpha_f$=1

\end{itemize}
We have this factor of $\frac{1}{2}$ in front of the achievable rates because we have two time slots of equal duration and $R_n$ , $R_f$  are the achievable rates during the first time slot alone.
\subsubsection{Relaying slot}
The next half of the time slot is called the relaying slot. As we saw, the near user already has the far user's data because he decoded it in the previous time slot. In the relaying time slot, the near user just transmits this data to the far user. The achievable rate of the far user at the end of the relaying slot is,
\begin{equation}
\mathbf{R}_{f,2}= \frac{1}{2} log_{2}  ( 1 + \rho \lvert \mathbf{h}_{nf}\lvert ^{2}) 
\end{equation}
here,  $h_{nf}$ is the channel between the near user and the far user. We can already see that $R_{f,2}>R_{f,1} $ because of two reasons:
\begin{itemize}
\item There is no interference from other transmissions
\item There is no fractional power allocation. The whole transmit power is given to the far user
\end{itemize}
\subsubsection{Diversity combining}
\begin{figure}[htbp]
\centerline{\includegraphics[width=8cm,height=8cm]{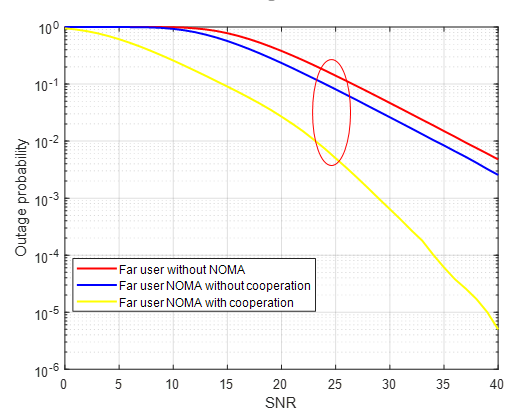}}
\caption{Outage probabilities of the considered transmission schemes: cooperative NOMA, non-cooperative NOMA, Orthogonal Multiple Access (OMA).}
\label{fig2}
\end{figure}

Now, at the end of the two-time slots, the far user has two copies of the same information received through two different channels. The far user can now use a diversity combining technique. For example, he can use selection combining to choose the copy which was received with high SNR. After selection combining, the achievable rate of the far user would be,
\begin{equation}\label{eu_eqn5}
\mathbf{R}_f= \frac{1}{2} log_{2}  ( 1 + \max (\frac{\alpha_f \rho \lvert \mathbf{h}_{f}\rvert^{2}}{{\alpha_{n}} \rho \lvert \mathbf{h}_{f}\rvert ^{2} + 1} ,\rho \lvert \mathbf{h}_{f}\rvert ^{2}) )
\end{equation}
If we do not  use cooperative relaying, the achievable rate of the far user would be,
\begin{equation}\label{eu_eqn9}
\mathbf{R}_{f,noncoop}= log_{2}  ( 1 + \frac{\alpha_{f} \rho \lvert \mathbf{h}_{f}\rvert ^2}{{\alpha_{n}} \rho \lvert \mathbf{h}_{f} \rvert ^{2} + 1}  )
\end{equation}
The factor of $\frac{1}{2}$ does not exist here because the entire time slot will be used for transmission in non-cooperative communication.
If we DID NOT use NOMA, for example, if we use Time-division multiple access (TDMA), we will allocate half of the time slot for transmission of the far user data. Hence, the achievable rate of the far user would be,
\begin{equation}\label{eu_eqn6}
\mathbf{R}_{f,OMA}= \frac{1}{2}log_{2}  ( 1 +  \rho \lvert \mathbf{h}_{f}\rvert ^{2} )
\end{equation}
Now that we have seen what cooperative communication is, and how it is useful to our network, let us design a cooperative NOMA network, that we order the performances of different schemes as cooperative NOMA $>$ non-cooperative NOMA $>$ Orthogonal Multiple Access (OMA). So, we can see from Fig.~\ref{fig2} that cooperative communication is beneficial.
\subsection{User Pairing in NOMA}
We have previously shown that NOMA can support numerous consumers in the same band simultaneously moment. The next logical step will be to figure out how many users can be supported per frequency carrier. In practice, if the number of users is increased beyond a certain point, the network's total throughput begins to decline. As a result, the number of users per carrier cannot be increased forever. 

\begin{figure}[htbp]
\centerline{\includegraphics[width=8cm,height=7cm]{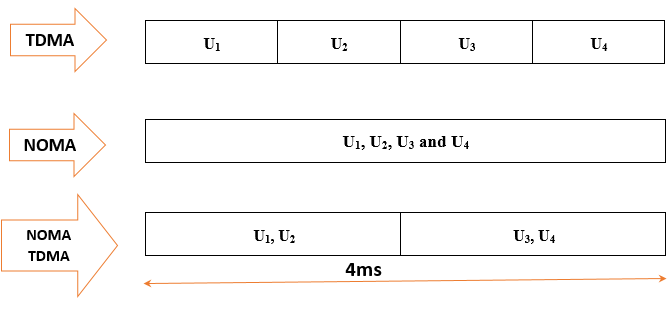}}
\caption{Hybrid NOMA.}
\label{fig3}
\end{figure}

A feasible option to support all users is to use hybrid NOMA. Hybrid NOMA is a technique that combines NOMA and any of the OMA approaches. Consider the combination of TDMA and NOMA, as indicated in Fig.\ref{fig3}. Assume we have a time slot with a duration of 4ms. Within this time frame, we must support four users. Now, TDMA will divide the 4 ms slot into four 1 ms slots, and each user will be assigned one slot. The four users will be given the entire 4 ms slot by NOMA. This, as we all know, will increase SIC complexity and processing time. In contrast, hybrid NOMA divides the 4 ms slot into two 2 ms slots and assigns two NOMA users to each slot. As can be shown, hybrid NOMA can provide service to all users while reducing complexity. As we progress through this article, we'll talk more about hybrid NOMA methods. Now we must figure out how to match users within each orthogonal resource that is offered. Should we group users together{(1,2),(3,4)} or {(1,3),(2,4)} or {(1,4),(2,3)}.
\section{Performance  Analysis}

Let us consider a downlink cooperative communication-based UAV scenario with four users $U_1, U_2, U_3 ~~and~ U_4$. Let $d_1, d_2, d_3 ~, and~ d_4$ denote the distances of those users, respectively, from the UAV.  $U_1$ is the nearest user and $U_4$ is the farthest user. Therefore, their channel conditions are ordered as $ \lvert \mathbf{h_1}\rvert^{2}  <  \lvert \mathbf{h_2} \rvert^{2} < \lvert\mathbf{h_3}\rvert^{2} < \lvert \mathbf{h_4}\rvert^{2}  $. We have two orthogonal resource blocks (time/ frequency/ subcarriers) to which two users must be allocated. We will pair users based on various distances. There are two basic ways to accomplish this:
\begin{itemize}
\item Near-far pairing (N-F)
\item Near-near, far-far pairing (N-N, F-F)

\end{itemize}
\subsubsection{Near-far pairing (N-F)} 
In this method, the nearest user to the UAV is paired with the farthest user from the UAV. The next nearest user is paired with the next farthest user and so on. In our example, $U_1$  is the nearest user and $U_4$ is the farthest user. So, N-F pairing will pair $U_1$ with $U_4$ in one resource block. $U_2$ will be paired with $U_3$ in the next resource block. 

In the first pair of users, $U_1$ is the near user and $U_4$ is the far user. Therefore, we have to choose the power allocation coefficients as $\alpha_1 <\alpha_4 $. So, $U_1$  should perform SIC, while $U_4$ will perform direct decoding. Similarly, in the second pair of users, $U_2$  is the near user and $U_3$  is the far user. Therefore, we have to choose $\alpha_2<\alpha_3 $. Here, $U_2$ should perform SIC while $U_3$ will perform direct decoding.

The achievable rates for the users in the first pair are,
\begin{equation}
\mathbf {R}_{1,nf}= \frac{1}{2} log_2  (1 + \frac {P\alpha_1 \lvert h_1\rvert ^2}{\sigma^2} )(after ~ SIC).
\end{equation}

\begin{equation}
\mathbf {R}_{4,nf}= \frac{1}{2} log_2  (1 + \frac {P\alpha_4 \lvert h_4\rvert ^2}{P\alpha_1 \lvert h_4\rvert ^2+\sigma^2 } ).
\end{equation}
Similarly, for the second pair,
\begin{equation}
\mathbf {R}_{2,nf}= \frac{1}{2} log_2  ( 1 + \frac {P\alpha_2 \lvert h_2\rvert ^2}{\sigma^2} )(after  ~ SIC).
\end{equation}

\begin{equation}
\mathbf {R}_{3,nf}= \frac{1}{2} log_2  ( 1 + \frac {P\alpha_3 \lvert h_3\rvert ^2}{P\alpha_2 \lvert h_3\rvert ^2+\sigma^2 } ).
\end{equation}
The sum rate of the (N-F) scheme will be:
\begin{equation}
\mathbf {R}_{nf}=\mathbf {R}_{1,nf}+\mathbf {R}_{2,nf}+\mathbf {R}_{3,nf}+\mathbf {R}_{4,nf}
.
\end{equation}

\subsubsection{Near-near, far-far pairing (N-N, F-F)} 
Another way to perform user pairing is to group the nearest user with the next nearest user. The farthest user is grouped with the next farthest user. If we follow this strategy, in our example, $U_1$ will be paired with $U_2$ in one resource block. $U_3$ will be paired with $U_4$ in the next resource block.

Now, in the first pair of users,  $U_1$ is nearest to the BS when compared to $U_2$. Therefore, we have to choose $\alpha_1 <\alpha_2 $ .  $U_1$ should perform SIC, $U_2$ will perform direct decoding. Similarly, $U_3$ is closer to the BS than $U_4$. So, we have to choose  $\alpha_3 <\alpha_4 $. $U_3$ should perform SIC, while $U_4$ will perform direct decoding.

The achievable rates for the users in the first pair are, 
\begin{equation}
\mathbf {R}_{1,nn}= \frac{1}{2} log_2  ( 1 + \frac {P\alpha_1 \lvert h_1\rvert ^2}{\sigma^2} )(after ~  SIC).
\end{equation}

\begin{equation}
\mathbf {R}_{2,nn}= \frac{1}{2} log_2  ( 1 + \frac {P\alpha_2 \lvert h_2\rvert ^2}{P\alpha_1 \lvert h_2\rvert ^2+\sigma^2 } ).
\end{equation}
Similarly, for the second pair,
\begin{equation}
\mathbf {R}_{3,nn}= \frac{1}{2} log_2  ( 1 + \frac {P\alpha_3 \lvert h_3\rvert ^2}{\sigma^2} )(after ~  SIC).
\end{equation}

\begin{equation}
\mathbf {R}_{4,nn}= \frac{1}{2} log_2  ( 1 + \frac {P\alpha_4 \lvert h_4\rvert ^2}{P\alpha_3 \lvert h_4\rvert ^2+\sigma^2 } ).
\end{equation}
The sum rate of the (N-F) scheme will be
\begin{equation}
\mathbf {R}_{nn}=\mathbf {R}_{1,nn}+\mathbf {R}_{2,nn}+\mathbf {R}_{3,nn}+\mathbf {R}_{4,nn}
.
\end{equation}
\section{Simulation Results}
We have observed two different AF relaying UAV cooperative NOMA approaches. The question is which is better, and what if we just multiplex all of the users on the same carrier (SC-NOMA), with no user pairing? Given the additional work required for user pairing, why do we need NOMA at all? Why not use TDMA instead? Let's find out by putting everything together in MATLAB. In this simulation, $N$ users with $N=4$ are considered. When we utilize each of the user pairing strategies with the UAV that we looked at, we will plot the network's sum rate. We will also examine the network's sum-rate performance with simply SC-NOMA and TDMA with UAV. 

\begin{figure}[htbp]
\centerline{\includegraphics[width=8cm,height=8cm]{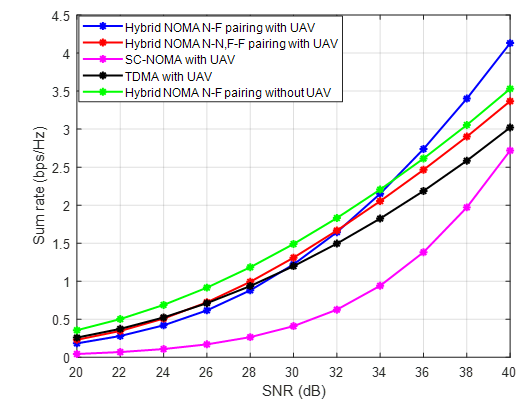}}
\caption{Sum-rate performance of the network with SC-NOMA, TDMA, hybrid NOMA N-F pairing and hybrid NOMA NN-FF pairing with/without UAV. }
\label{fig4}
\end{figure}
When the near user is coupled with the far user, as seen in Fig.\ref{fig4}, a higher sum rate is produced. This supports the well-known notion that cooperative NOMA operates better when the channel conditions, or the LoS probability model in UAV communication, are different between the two users. When near-near, far-far pairing is employed, NOMA-based UAV still outperforms TDMA-based UAV in terms of sum rate, but the difference is not significant. When compared to TDMA, the performance of SC-NOMA-based UAVs is poor since all users are crammed onto the same carrier, generating interference. This also validates our assumption that increasing the number of customers sharing the same carrier without paying a premium is impossible. Another benefit of the N-F pairing UAV relay over the N-N-F-F pairing UAV relay is the careful selection of users who may be impacted by interference. In contrast to the N-N-F-F system, which gives interference-free transmission to one near the user and one far user (assuming perfect SIC), the N-F strategy provides interference-free reception to both near users (assuming perfect SIC). As a result, N-F has a higher sum rate. When noise dominates the interference power, however, the choice of users who experience interference makes a little effect. Both systems are projected to perform similarly.

\section{Conclusion and Future Work  }

In this paper, a cooperative NOMA user pairing under an LoS channel is proposed, which is provided by a UAV serving as an AF relay. We investigated two user pairing strategies to improve downlink transmission performance. According to our results, implementing a user pairing strategy offers the most appropriate use of power and bandwidth resources. If we couple users at random, we may not be able to maximize the network's potential. The conversion of our single-antenna UAV relay model to a multiple-antenna UAV relay should be interesting. These issues will be addressed in the future.

\vspace{12pt}
\color{red}


\begin{thebibliography}{00}

\bibitem{b1} C. Zhang and W. Zhang, “Spectrum sharing for drone networks,”
IEEE J. Sel. Areas Commun., vol. 35, no. 1, pp. 136–144, Jan. 2017.
\bibitem{b2} J. Ji, K. Zhu, D. Niyato, and R. Wang, “Joint cache placement, flighttrajectory, and transmission power optimization for multi-UAV assistedwireless networks,” IEEE Trans. Wireless Commun., vol. 19, no. 8,pp. 5389–5403, Aug. 2020.
\bibitem{b3} S. Zhang, H. Zhang, B. Di, and L. Song, “Cellular UAV-to-X communications:
Design and optimization for multi-UAV networks,” IEEE
Trans. Wireless Commun., vol. 18, no. 2, pp. 1346–1359, Feb. 2019.
\bibitem{b4} Z. Wang, L. Duan, and R. Zhang, “Adaptive deployment for UAVaidedcommunication networks,” IEEE Trans. Wireless Commun.,
vol. 18, no. 9, pp. 4531–4543, Sep. 2019.
\bibitem{b5} Y. Zeng, J. Xu, and R. Zhang, “Energy minimization for wireless communication
with rotary-wing UAV,” IEEE Trans. Wireless Commun.,
vol. 18, no. 4, pp. 2329–2345, Apr. 2019.
\bibitem{b6} M. Mozaffari, W. Saad, M. Bennis, and M. Debbah, “Unmanned aerialvehicle with underlaid device-to-device communications: Performance and tradeoffs,” IEEE Trans. Wireless Commun., vol. 15, no. 6, pp. 3949–3963, Jun. 2016.
\bibitem{b7} X. Liu, J. Xu, and H. Tang, “Analysis of frequency-dependent line-of-sight probability in 3-D environment,” IEEE Commun. Lett., vol. 22, no. 8, pp. 1732–1735, Aug. 2018.
\bibitem{b8} M. Mozaffari, W. Saad, M. Bennis, and M. Debbah, “Mobile
unmanned aerial vehicles (UAVs) for energy-efficient Internet of things
communications,” IEEE Trans. Wireless Commun., vol. 16, no. 11,
pp. 7574–7589, Nov. 2017.
\bibitem{b9} C. Zhan, Y. Zeng, and R. Zhang, “Energy-efficient data collection in UAV enabled wireless sensor network,” IEEE Wireless Commun.
Lett., vol. 7, no. 3, pp. 328–331, Jun. 2018.
\bibitem{b10} Y. Zeng and R. Zhang, “Energy-efficient UAV communication with
trajectory optimization,” IEEE Trans. Wireless Commun., vol. 16,
no. 6, pp. 3747–3760, Jun. 2017.

\bibitem{b11} A. Farajzadeh, O. Ercetin, and H. Yanikomeroglu, “UAV data collection
over NOMA backscatter networks: UAV altitude and trajectory
optimization,” in Proc. IEEE Int. Conf. Commun. (ICC), Shanghai,
China, May 2019, pp. 1–7.
\bibitem{b12} Z. Wei, L. Yang, D. W. K. Ng, J. Yuan, and L. Hanzo, “On the
performance gain of NOMA over OMA in uplink communication
systems,” IEEE Trans. Commun., vol. 68, no. 1, pp. 536–568, Jan.
2020.
\bibitem{b13} Z. Ding et al., “Impact of user pairing on 5G nonorthogonal multipleaccess
downlink transmissions,” IEEE Trans. Veh. Technol., vol. 65,
no. 8, pp. 6010–6023, Aug. 2016.
\bibitem{b14} M. Shirvanimoghaddam et al., “Massive non-orthogonal multiple access
for cellular IoT: Potentials and limitations,” IEEE Commun. Mag.,
vol. 55, no. 9, pp. 55–61, Sep. 2017.

\bibitem{b15}Y. Liu, Z. Qin, M. Elkashlan, Z. Ding, A. Nallanathan, and L. Hanzo,
“Nonorthogonal multiple access for 5G and beyond,” Proc. IEEE,
vol. 105, no. 12, pp. 2347–2381, Dec. 2017.
\bibitem{b16} S. M. R. Islam, M. Zeng, O. A. Dobre, and K.-S. Kwak, “Resource
allocation for downlink NOMA systems: Key techniques and open
issues,” IEEE Wireless Commun., vol. 25, no. 2, pp. 40–47, Apr. 2018.
 \bibitem{b17}S. M. R. Islam, N. Avazov, O. A. Dobre, and K.-S. Kwak, “Powerdomain
non-orthogonal multiple access (NOMA) in 5G systems: Potentials
and challenges,” IEEE Commun. Surveys Tuts., vol. 19, no. 2,
pp. 721–742, 2nd Quart., 2017.
\bibitem{b18}W. Ni, X. Liu, Y. Liu, H. Tian and Y. Chen, "Resource Allocation for Multi-Cell IRS-Aided NOMA Networks," in IEEE Transactions on Wireless Communications, vol. 20, no. 7, pp. 4253-4268, July 2021.
\bibitem{b19}  N. Iswarya,  L.S.Jayashree, "A Survey on Successive Interference Cancellation Schemes in Non-Orthogonal Multiple Access for Future Radio Access". Wireless Pers Commun 120, 1057–1078 ,2021.
\bibitem{b20} J. Won, D. Kim, Y.Park, J. Lee,
"A survey on UAV placement and trajectory optimization in communication networks: From the perspective of air-to-ground channel models," ICT Express, 2022.

\bibitem{b21} Ajmal, Mahnoor, and Muhammad Zeeshan. "A Novel Hybrid AF/DF Cooperative Communication Scheme for Power Domain NOMA." 2021 IEEE 15th International Symposium on Applied Computational Intelligence and Informatics (SACI). IEEE, 2021.
\bibitem{b22}V.N. Nayak,K.K. Gurrala, K.K. "Enhanced Physical Layer Security for Cooperative NOMA Network with Hybrid-Decode-Amplify-Forward Relaying via Power Allocation Assisted Control Jamming". Wireless Pers Commun 120, 2473–2490 2021.



\end{thebibliography}
\end{document}